\documentclass[twocolumn,preprintnumbers,amsmath,amssymb]{revtex4}

\usepackage{graphicx}  
\usepackage{epstopdf}
\begin{document}

\title{
Two-dimensional plasmons 
in lateral  carbon nanotube network structures 
and\\ their effect on the terahertz radiation  detection
}
\author{V. Ryzhii$^{1,2,3}$,  T. Otsuji$^1$, M. Ryzhii$^4$, 
V. G. Leiman$^5$, G. Fedorov$^{5,6}$,
G. N. Goltzman$^6$,\\ I. A. Gayduchenko$^{6,7}$, N. Titova$^6$,
 D. Coquillat$^8$,
 D. But$^8$, W. Knap$^8$,\\
 V. Mitin$^9$, and  M. S. Shur$^{10}$.
 }
\affiliation{$^1$ Research Institute of Electrical Communication, Tohoku University,  Sendai 980-8577, Japan\\ 
$^2$ Institute of Ultra High Frequency Semiconductor Electronics of RAS, Moscow 117105, Russia\\
$^3$ Research Center of Photonics and Infrared Engineering, Bauman Moscow State Technical University, Moscow 111005, Russia\\
$^4$ Department of Computer Science and Engineering, University of Aizu, Aizu-Wakamatsu 965-8580, Japan\\
$^5$ Department of General Physics, Moscow Institute of Physics and Technology, Dolgoprudny, 147100, Russia\\
$^6$ Physics Department, Moscow State Pedagogical University, Moscow 119991, 
Russia\\
$^7$ National Research Center "Kurchatov Institute," Moscow 123182, Russia\\
$^8$ Laboratoire Charles Coulomb UMR 5221, Universite Montpellier 2 and CNRS, F-34095, Montpellier, France\\ 
$^9$ Department of Electrical Engineering, University at Buffalo, Buffalo, NY 1460-1920, USA\\ 
$^{10}$ Departments of Electrical, Computer, and Systems Engineering and Physics, Applied Physics, and Astronomy, Rensselaer Polytechnic Institute, Troy, NY 12180, USA
}

 \begin{abstract} 
{\bf Keywords:} carbon nanotube network, Schottky contact, two-dimensional carrier system, terahertz radiation,  plasmonic resonance
 
 We consider the carrier transport and  plasmonic phenomena 
 in the
 lateral carbon nanotube (CNT) networks forming the device channel
 with asymmetric electrodes. One electrode is the Ohmic contact to the CNT network and the another contact is the Schottky contact.
 These structures can  serve as detectors of the terahertz (THz) radiation.
 We develop the device model for response of  the lateral CNT networks which comprise a mixture of randomly oriented semiconductor CNTs (s-CNTs) and quasi-metal CNTs ( m-CNTs).  
 The proposed model includes the concept of the two-dimensional  plasmons in relatively dense networks of randomly oriented CNTs (CNT "felt") and predicts the detector responsivity spectral characteristics. 
 The detection mechanism is the rectification of the ac current due the nonlinearity of the Schottky contact current-voltage characteristics under the conditions of a strong enhancement of the potential drop at this contact associated with the plasmon excitation.
 We demonstrate that  the excitation of the two-dimensional plasmons  by incoming THz radiation the detector responsivity  can 
induce sharp resonant peaks of the detector responsivity at the signal frequencies
corresponding to the plasmonic resonances. 
The detector responsivity depends on the fractions of the s-
and m-CNTs. The burning of the near-contact regions of the m-CNTs or destruction  of these CNTs  leads to a marked increase in the responsivity in agreement with our experimental data.
The  resonant THz  detectors with sufficiently dense lateral CNT networks can compete and  surpass other THz detectors using plasmonic effects at room temperatures.
\end{abstract} 
 
\maketitle

\newpage
\section{Introduction}
Plasmonic electron phenomena in semiconductors could enable advanced terahertz (THz) devices~\cite{1,2}. In particular,
the THz detectors
 using the non-linear plasmonic properties of  two-dimensional electron gas (2DEGs)~\cite{2} have proved to be very effective devices operating at room temperature
\cite{3,4,5,6,7,8,9,10,11,12,13,14,15,16,17,18,19,20,21}. 
The resonant excitation of the plasma oscillations in the gated and ungated 2DEGs by incoming THz radiation could support  relatively large
 electric-field oscillations and, hence, rather large rectified current components. Using the nonlinearity of the current-voltage characteristics (such as the nonlinearity of the thermionic or tunneling channel-gate leakage or injection currents) 
 to enhance the hydrodynamic nonlinearity of plasma oscillations~\cite{2,22,23,24,25,26,27,28,29},
the pertinent detectors can exhibit marked advantages. In particular, 
the combination of the plasmonic properties of the 2DEG channel with
the contact nonlinearity can be realized in
lateral Schottky diodes
\cite{24,25}. 
 
 The detection of THz radiation in lateral device structures based on
 carbon nanotubes (CNTs)  was demonstrated in Refs.~\cite{30,31,32,33,34}. The detection mechanisms in the devices with nonuniform CNT networks~\cite{32,33,34} are associated with nonuniform
 heating of the electron or hole system by the absorbed THz radiation or with 
the structure asymmetry due to metal contacts with different nonlinear properties leading to
the rectification of the ac current induced by this radiation.
Different CNT-base photodetectors and mechanisms of their operation were reviewed
in Ref.~\cite{35}
The incident THz radiation can excite the plasma oscillations in
lateral CNT networks (with the wavelength larger that the average distance between the CNTs) in a similar way as in the 2DEG channels in the standard heterostructures or in the graphene-based heterostructures.
The plasmonic nature of the THz response in CNT systems was discussed in a number of publications (see, for example, Refs.~\cite{36,37,38,39,40}). The role of the plasmonic effects in
the  aligned CNTs and disordered CNT networks was experimentally confirmed in  Refs.~\cite{41,42,43,44}.
The interpretation of  the experimental findings, relied on the concept of 1D plasmons propagating
along the CNT and having cylindrical symmetry of the potential distribution
 (see, for example, Ref.~\cite{37}). However, in sufficiently dense lateral CNT,  different CNTs  interact with each other and,  as a result, the plasma oscillations excited by the THz radiation can be associated with the {\it collective coherent} motion of the carriers belonging to many CNT responding to
the  ac {\it self-consistent} electric fields. This collective motion  reveals the 2D-nature of the response, especially pronounced
 when the plasma wavelength markedly exceeds the average distance between the CNTs.

In this paper, in contrast to the previous works focused on the dilute CNT systems (including
the composite materials with randomly dispersed, randomly oriented CNTs), 
  we study  the 2D plasma oscillations excited by the THz radiation
in lateral {\it dense} CNT disordered networks consisting of a mixture of the single-wall semiconducting CNTs (s-CNTs) and quasi-metallic CNTs (m-CNTs) of the p-type.Also, we consider the networks
of relatively long CNTs connecting two highly conducting contacts with  distinct
properties and focus on the role
of the hole collective motion, i.e., on the plasmonic effects on the  THz radiation detection and
the characteristics of  THz detectors based on such CNT networks.

The CNT device structure with relatively dense CNT network under consideration is similar to that fabricated 
and studied experimentally recently (type B devices for sub-THz and THz detection ~\cite{34} and for transistor applications~\cite{45}).
We demonstrate that the hole system in the lateral CNT network can be treated as a 2D hole gas (2DHG), exhibiting  pronounced plasmonic properties, and that the excitation of the plasma oscillations can result in a resonant enhancement
rectified current component. Such CNT-based structures can be used in uncooled CNT-detectors with elevated responsivity
exploiting the rectification mechanism.

\section{Device model}

The device structure under consideration  is shown in Fig.~1.
We assume that the right-hand side  metal (say, palladium) electrode  to the lateral CNT network forms the Schottky contact (for the s-CNTs), whereas the left contact (vanadium) is virtually Ohmic. The structure is doped with acceptors. The highly conducting substrate (doped Si), separated from the CNT network by a barrier layer
(SiO$_2$),
plays the role of the gate. The variation of the gate voltage $V_g$ provides a variation of the population of both s- and m-CNTs with carriers (holes) and a change in the average sheet
charge density in the CNT network. 
Figure~2 shows the band diagram of the CNT structure under consideration
at a negative  gate voltage.
The CNT array is attracted to the SiO$_2$ layer
by the Van-der-Waals forces.
The CNT network (CNT "felt") consists of randomly oriented sufficiently long CNTs (${\cal L}\gg L$, where ${\cal L}$ and $L$ are the characteristic length of the CNTs and the spacing between the contacts, respectively), so that even those CNTs directed
at rather large angles $\alpha$ (close to $\pm\pi/2$) to the direction of the source-drain current  are contacted both
the electrodes.
The device is coupled to the THz radiation with an antenna which also serves
to provide the dc and ac (induced by the incoming THz radiation)  voltages between the contacts. 

\begin{figure}[t]
\centering
\includegraphics[width=6.0cm]{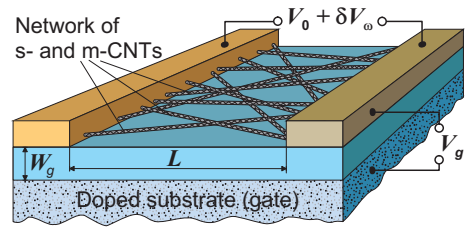}
\caption{Schematic view of   CNT device  structure.
}
\end{figure}

\begin{figure}[t]
\centering
\includegraphics[width=6.0cm]{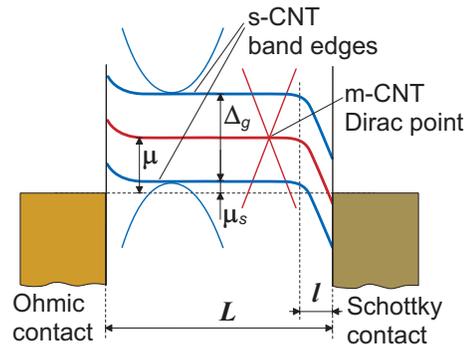}
\caption{ Schematic view  of s- and m-CNTs band diagrams in the  structure with Ohmic (left) and Schottky (right) contacts  at $V_g < 0$.
}
\end{figure}

The ratio of the densities (per unit length in the in-plain direction parallel to the contacts)of the CNTs and m-CNTs,
$N_m$ and $N_s$, can vary in a wide range starting from $N_m/N_s = 1/2$
to zero. 
The latter case corresponds to the situation when the m-CNTs are selectively destructed by applying large enough source-drain voltage at positive gate voltage with s-CNTs in non-conducting state.

For simplicity, the carrier (hole) dispersion in the s-CNTs is assumed to have a one-dimensional parabolic energy dependence on the momentum $p$ with the effective mass $m_s$: $\varepsilon = \Delta_g/2 + p^2/2m_s$
The dispersion of m-CNTs can be described by the quasi-relativistic dispersion relation which in the case of small energy gap can be presented as $\varepsilon \simeq v_Wp$. Here $\Delta_g$ is the energy gap in the s-CNTs and $v_W \simeq 10^8$~cm/s is the characteristic velocity of the gapless spectrum~\cite{46}. We restrict our consideration to sufficiently dense  CNT networks
with a large number, ${\cal N}$, of the holes in the area with  the radius equal to 
the Debye  
screening length $l_D$. In this situation, the lateral carrier movement is associated with the effect of the self-consistent  electric field.
However, the average distance between the CNTs $N_m^{-1}, N_s^{-1}$ is assumed
to be much larger than the size of the  localization region  around individual
CNTs, so that the wave function overlap is insignificant. This allows us to consider the carrier system in each CNT as one-dimensional. The pertinent estimates will be provided in the following. 

The net voltage between the source and drain contacts is equal to $V = V_0 +\delta V_{\omega}\exp(-i\omega\,t)$, i.e., it comprises the dc bias voltage $V_0$
and the ac THz signal voltage $\delta V_{\omega}\exp(-i\omega\,t)$ with the amplitude $\delta V_{\omega}$ and frequency $\omega$

In the presence of the ac electric field $\delta E_{\omega} = - d\delta \varphi_{\omega}/dx$, where $\delta\varphi_{\omega}$
is the ac component of the electric potential at the CNT network plane,
the ac component of hole distribution function $\delta f_{\omega}(p,\alpha)$,
where $p$ is the hole momentum along the CNT 
and $\alpha$ is the angle between the source-drain current direction and the CNT axis, can be found from the kinetic (Boltzmann) equation:

\begin{equation}\label{1}
(i\omega + \nu)\delta f_{\omega}(p,\alpha) = -e\cos\alpha\frac{d\delta \varphi_{\omega}}{dx}
\frac{d f_0(p)}{d p}.
\end{equation}
 Here $\nu$ is the phenomenological electron scattering frequency due to
 the electron scattering with disorder ($\nu = \nu_s$ and $\nu = \nu_m$ in the semiconducting and quasi-metal CNTs, respectively, $f_0(p)$ is the equilibrium Fermi
 distribution function, and $e = |e|$ is the electron charge.
 In sufficiently dense CNT networks, the intersections of the CNTs can occur.
 In the model under consideration, these intersections are accounted for as additional scattering points for the holes propagating along the CNTs leading to an increase of the frequencies $\nu_m$ and $\nu_s$. 
 
\subsection{Semiconducting CNT network} 
 
Generally both the electron motion along CNTs and and their polarization contribute
to the ac conductivity of the CNT network. The polarization component is small because of a large energy of the bound states associated with the direction perpendicular to the CNT axis except the signal frequency close to the polarization resonant frequency $\omega_{\perp}$~\cite{38}. As shown for the quantum wires (QWRs) of different origin~\cite{38}, $\omega_{\perp} \propto 1/\sqrt{R_{CNT}}$, where $R_{CNT}$ the QWR
radius. In the case of CNTs with $2R_{CNT} \simeq 1.5 - 2.3$~nm, the resonant frequency
$\omega_{\perp}$ is markedly large, so that the pertinent resonances can play a substantial role in the range of  rather high signal frequencies which is out of scope of this work.

In the framework of a simplified model , the ac in-plane conductivity of the s-CNT network  averaged over the CNT angles $\langle \sigma_{\omega}\rangle$ can be presented as

\begin{multline}\label{2}
\langle \sigma_{\omega}^s\rangle = \frac{2eN_s}{\pi\hbar}\int_{-\pi/2}^{\pi/2}d\alpha\,\cos(\alpha)\, \Theta(\alpha)
\int_{-\infty}^{\infty}\frac{dpp}{m_s}\,\delta f(p,\alpha)\\
=
\frac{2ie^2N_s}{\pi\hbar\,m_s}\int_{-\pi/2}^{\pi/2} d\alpha\,\cos^2(\alpha)\,\Theta(\alpha)\\
\times\int_{-\infty}^{\infty}\frac{dpp}{(\omega + i\nu_s)}\biggl[-\frac{d f_0(p)}{d p}
\biggr]\\
= \frac{i4\sqrt{2}e^2N_s\theta}{\pi\hbar\sqrt{m_s}}\int_{0}^{\infty}\frac{d\varepsilon\sqrt{\varepsilon}}{(\omega + i\nu_s)}\biggl[-\frac{d f_0(\varepsilon)}
{d \varepsilon}
\biggr].
\end{multline}
Here  $\hbar$ is the reduced Planck constant, $\Theta(\alpha)$ is the distribution function  of the CNTs over their direction angle $\alpha$, and
$\theta = \int_{-\pi/2}^{\pi/2}d\alpha\cos^2(\alpha)\,\Theta(\alpha)$.
The angle distribution parameter $\theta = 1$ and $\theta = 1/2$ in the cases when all the CNTs are perpendicular to the contact and when they are uniformly distributed over the angles, respectively. 

Assuming that the one-dimensional hole system in each CNT is degenerate, i.e.,
(the electron Fermi energy in the semiconducting CNT $\mu_s = \mu -\Delta_g/2$ markedly exceeds the temperature $T$(in the energy units). The Fermi energy $\mu$ is counted from the middle of the 
the conducting CNTs energy gap, i,e,m from the Dirac point.
From Eq.~(2) we obtain 

\begin{equation}\label{3}
\langle\sigma_{s,\omega}\rangle \simeq \frac{i4e^2N_s\theta}{\pi\hbar}
\sqrt{\frac{2\mu_s}{m_s}}\frac{1}{(\omega + i\nu_s)}.
\end{equation}
Considering that the dc surface hole density  in the semiconducting
CNTs$\Sigma_0^s$
is given by the following formula:

\begin{equation}\label{4}
\Sigma_{s,0} = \frac{4N_s}{2\pi\hbar}\int_{-\infty}^{\infty}dp f_0(p) \simeq 
\frac{4N_s\sqrt{2m_s\mu_s}}{\pi\hbar},
\end{equation}
Eq.~(3) can be presented as

\begin{equation}\label{5}
\langle\sigma_{s,\omega}\rangle \simeq \frac{ie^2\theta\Sigma_{s,0}}
{m_s(\omega + i\nu_s)}.
\end{equation}

The appearance of a factor of $\theta$  ($1/2 \leq \theta \leq 1$) in Eq.~(5) is associated with the angle spreading of the CNTs and a decreased contribution of the CNTs not parallel to the current direction.  

\subsection{Quasi-metallic CNT network}

The ac conductivity associated with the quasi-metallic CNTs of the density $N_m$ can be calculated using the following expression:

\begin{multline}\label{6}
\langle \sigma_{m,\omega}\rangle = \frac{2eN_mv_W}{\pi\hbar}\int_{-\pi/2}^{\pi/2}
d\alpha\,\cos(\alpha)\,\Theta(\alpha)
\int_{-\infty}^{\infty}dp\,\delta f(p,\alpha)\\
 = \frac{i4e^2N_m\theta}{\pi\hbar}\frac{v_W}{(\omega + i\nu)}\frac{1}
 {[1 + \exp(-\mu_m/T)]}\\
 \simeq \frac{i4e^2N_m\theta}{\pi\hbar}\frac{v_W}{(\omega + i\nu_m)},
\end{multline}
Here $\mu_m = \mu \gg T$. 
Simultaneously for the relation between the hole density in the quasi-metallic CNTs $\Sigma_{m,0}$ and the Fermi energy $\mu_m$ and between the ac conductivity
and the hole density one can obtain

\begin{equation}\label{7}
\Sigma_{m,0} \simeq \frac{4N_m\mu_m}{\pi\hbar\,v_W},
\qquad
 \langle\sigma_{m,\omega}\rangle \simeq \frac{ie^2\theta\Sigma_{m,0}}
 {m_m(\omega + i\nu_m)}.
\end{equation}
  The quantity $m_m = \mu_m/v_W^2 = \mu/v_W^2$ in degenerate electron systems with a linear dispersion is usually called as the fictituous mass.   

\subsection{Comparison of the ac conductivities}

As follows from Eq.~(6), the ac conductivity
$\langle \sigma_{m,\omega}\rangle$ does not explicitly depend on the Fermi energy
$\mu_s$

However, $\langle \sigma_{m,\omega}\rangle$ (as well as $\langle \sigma_{s,\omega}\rangle$)  can depend on the Fermi energy and the hole density via the dependences of the collision frequencies. 

Comparing Eqs.~(3) and (6), we find 

\begin{equation}\label{8}
\frac{\langle \sigma_{s,\omega}\rangle}{\langle \sigma_{m,\omega}\rangle}
= 
\frac{\sqrt{2\mu_s/m_s}}{v_W}\biggl(\frac{\omega + i \nu_m}{\omega + i \nu_s} \biggr)\frac{N_s}{N_m}
\end{equation}

If the collision frequency is proportional to the density of states at the Fermi
level, Eq.~(8)
 in the limits $\omega \ll \nu_s, \nu_m$ and $\omega \gg \nu_s, \nu_m$ respectively yields

\begin{equation}\label{9}
\frac{\langle \sigma_{s,0}\rangle}{\langle \sigma_{m,0}\rangle}
= 
\frac{\sqrt{2\mu_s/m_s}}{v_W}\biggl(\frac{ \nu_m}{ \nu_s} \biggr)\frac{N_s}{N_m} \simeq \frac{2\mu_s}{m_sv_W^2}\frac{N_s}{N_m}
\end{equation}
and

\begin{equation}\label{10}
\frac{\langle \sigma_{s,\infty}\rangle}{\langle \sigma_{m,\infty}\rangle}
= 
\frac{N_s}{N_m}\frac{\sqrt{2\mu_s/m_s}}{v_W}.
\end{equation}
Setting $N_s/N_m = 2$, $m_s = 6\times 10^{-29}$~g, and $\mu_s = 25 - 50$~meV,
we obtain $\langle \sigma_{0}^s\rangle/\langle \sigma_{0}^m\rangle \simeq
0.27 - 0.54 $ and $\langle \sigma_{s,\infty}\rangle/\langle \sigma_{m,\infty}\rangle \simeq 0.73 - 1.03$.

\section{Forced plasma oscillations in CNT network structures} 

\subsection{Gated structures}

For the CNT structures with the the gate layer thickness $W_g \ll L$,
the self-consistent ac potential at the CNT network plane $\delta \varphi_{\omega} = \delta\varphi_{\omega}(x) = \delta\psi(x, z)_{\omega}|_{z = 0} $ can be calculated from
the Poisson equation in the gradual channel approximation~\cite{47}:

\begin{equation}\label{11}
\frac{\delta\varphi_{\omega}}{W_g} = \frac{4\pi e}{\kappa}\delta\Sigma_{\omega}.
\end{equation}
Here $\delta \Sigma_{\omega} = \delta \Sigma_{s,\omega} + 
\delta \Sigma_{m,\omega}$ is the ac variation of the net hole
density in the CNT network,  $\kappa$ is the  dielectric constant of the gate layer  material. The axis $x$ is directed in the CNT network plane from one the contact to 
another and the axis $z$ is perpendicular to this plane (see Fig.1).

In Eq.~(11), the quantum capacitance~\cite{48} is disregarded because for the practical values of the gate layer thickness~\cite{33} 
the geometrical capacitance per unit area $C_g =\kappa/4\pi\,W_g$ is much smaller than the quantum capacitance.

The continuity equations of the electron components in the semiconducting and
quasi-metallic CNTs can be presented in the following form:

\begin{equation}\label{12}
-i\omega e\delta\Sigma_{s,\omega} + \langle \sigma_{s,\omega}\rangle
\frac{d^2 \delta\varphi_{\omega}}{d x^2}= 0,
\end{equation}
 
\begin{equation}\label{13}
-i\omega e\delta\Sigma_{m,\omega} + \langle \sigma_{m,\omega}
\rangle\frac{d^2 \delta\varphi_{\omega}}{d x^2}= 0.
\end{equation}
Combining Eqs.~(11) - (13), we arrive at the following equation governing the
ac potential in the channel

\begin{equation}\label{14}
\frac{4\pi\,W_g\langle \sigma_{\omega}\rangle}{i\omega\kappa}\frac{d^2 \delta\varphi_{\omega}}{d x^2} + \delta\varphi_{\omega}= 0,
\end{equation}
where $\langle \sigma_{\omega}\rangle = \langle \sigma_{s,\omega}\rangle + 
\langle \sigma_{m,\omega}\rangle$,
or

\begin{equation}\label{15}
\frac{4\pi\,W_g}{\kappa\omega}\biggr(\frac{D_s}{\omega + i\nu_s} + \frac{D_m}{\omega + i\nu_m} \biggr)\frac{d^2 \delta\varphi_{\omega}}{d x^2} + \delta\varphi_{\omega}= 0.
\end{equation}
Here $$
D_s = \frac{4e^2N_s\theta}{\pi\hbar}
\sqrt{\frac{2\mu_s}{m_s}} = \frac{e^2\theta\Sigma_{s,0}}{m_s}, 
$$
$$
  D_m = \frac{4e^2N_mv_W\theta}{\pi\hbar} =\frac{e^2\theta\Sigma_{m,0}}{m_m} .
$$
Assuming for simplicity that $\nu_s \simeq \nu_m = \nu$, Eqs.~(14) and (15) can be reduced to the following:

\begin{equation}\label{16}
\frac{d^2 \delta\varphi_{\omega}}{d x^2} + \ae^2_{\omega}\delta\varphi_{\omega}= 0.
\end{equation}
Here 
\begin{equation}\label{17}
\ae_{\omega} = \frac{\sqrt{\omega(\omega + i\nu)}}{s} = \frac{\pi\sqrt{\omega(\omega + i\nu)}}{2\Omega\,L},
\end{equation}
where

\begin{equation}\label{18}
s = \sqrt{4\pi\,W_g(D_s + D_m)/\kappa}
\end{equation}
 and

\begin{equation}\label{19}
\Omega_g = \frac{\pi\,s}{2L}
\end{equation}
are  the characteristic plasma-wave velocity and frequency of plasma oscillations in the gated channel of the length $L$ ($L \gg l$).
If $D_m \ll D_s$ and $\theta = 1$, the latter equation yield the same value for $s$ as for
the 2DEG with the parabolic electron spectrum and the effective mass $m$: $s =\sqrt{4\pi\,e^2\Sigma_{s,0}W_g/m\kappa}$.

Considering Eqs.~(4), (7), and (18), $\Omega_g$ can be expressed either via $\Sigma_{s,0}/m_s$
and $\Sigma_{m,0}/m_m$ or via $\mu$:

\begin{multline}\label{20}
\Omega_g = \frac{\pi}{2L}
\sqrt{\frac{4\pi\,e^2\theta\,W_g}{\kappa}
\biggl(\frac{\Sigma_{s,0}}{m_s} + \frac{\Sigma_{m,0}}{m_m}\biggr)}\\
 = 
\frac{\pi}{2L}\sqrt{\frac{4\pi\,e^2\theta\,W_g
(\Sigma_{s,0} + \overline{\Sigma_{m,0}})}{\kappa\,m_s}}. 
\end{multline}
%
Here $\overline{\Sigma_{m,0}} = (4m_sv_WN_m/\pi\hbar)$. This quantity does not depend
on the hole density and, hence, on the absolute value of the gate voltage $|V_g|$, and increase in the  
hole density leads to an increase in the fictituous mass $m_m$, so that
$\Sigma_{m,0}/m_m = const$ [see Eq.~(7)]. At $N_m = 2\times (10^4 - 10^5)$~cm$^{-1}$,
one obtains $\overline{\Sigma_{m,0}} \simeq (10^{11} - 10^{12})$~cm$^{-2}$.
It is instructive that $\overline{\Sigma_0^m}$ can be both smaller and larger
than real hole density in the m-CNTs depending the m-CNT density.
Since the net hole density (and, what is important, the hole density in the s-CNTs) in the gated structures varies with varying  gate voltage $V_g$,  the characteristic plasma
frequency $\Omega_g$ is voltage controlled. 
 However, the
 $\Omega_g - V_g$ dependence differ from those in the gated 2D electron or hole systems in the standard heterostructures ($s \propto V_g^{1/2}$~\cite{2}) and graphene ($s \propto V_g^{1/4}$~\cite{49}), because according to Eq.~(21)only the variations of the hole density in the s-CNTs affect the plasma frequency.

The boundary conditions can be formulated using the following reasoning.
The resistance of the left-hand-side contact (at $x = 0$), which can be considered as the Ohmic one, is small so that the potential drop at this contact is negligible.
In contrast, the potential drop across the right-hand-site contact depletion region, which is usually  modeled as
a Schottky contact~(see, for example, Refs. \cite{50,51} ).

In our model, we assume that the ac current density across the depletion region
is associated with the Ohmic component  (characterized by the m-CNT contact resistance $R_m$), hence, the pertinent current density 
$\delta j_m = \delta\varphi_{\omega}N_m/R_m$,  
with the current density through the contacts of the s-CNTs and metallic electrode (which is modelled by the Schottky-contact current-voltage  characteistic)
$\delta j_s \simeq N_s(\delta \varphi_{\omega}/R_s  + 
\beta^s\delta\varphi_{\omega}^2)|_{x = L-l}$, 
and the density of the displacement current  
$\delta^{disp} = - i\omega\,C$.
Here $R_s = (dJ_s/d\varphi)^{-1}$ is the resistance of the  Schottky contact
between each s-CNT and the electrode, 
$\beta_s = \frac{1}{2}(d^2J_s/d\varphi)^2$ is the parameter of the Schottky-contact nonlinearity,  $C$ is the capacitance of the depletion region (per unit length in the lateral direction perpendicular to the current flow), and  $l$ is the thickness of the depletion region at the Schottky contact
[see Fig.~2]. 
Thus, the net current density $\delta j_C $ across the contact is equal to

\begin{equation}\label{21}
\delta j_C \simeq Y_{\omega}\delta\varphi_{\omega}|_{x = L-l}
 + N_s\beta_s\delta\varphi_{\omega}^2|_{x = L-l},
\end{equation}
where 
$Y_{\omega} = 1/r - i\omega\,C$ is the contact region admittance (per unit length
 the current flow direction)
and $1/r = N_m/R_m + N_s/R_s$, so that $Y_0 = 1/r$.
In the case of the blade-like electrodes,     the capacitance $C$ can be estimated
as $C <  (\kappa/2\pi^2)\Lambda$, where the factor
$\Lambda \sim 1 $   describes the specifics of the geometry the Schottky contact
of the s-CNTs and the metal electrodes. )n the analogy with the Schottky contacts
of 2DEG or 2DHG, $\Lambda \simeq \kappa/2\pi^2\ln (4L/l)$~\cite{52} (see also Refs.~\cite{53,54,55}. In the gated structures, the net ac current between the electrodes comprises also the displacement current between them through the gate
$\delta j_G = -i\omega C_{sgd}\delta V_{\omega}$. The pertinent capacitance (per unit length)
can be estimated as $C_{sgd} \simeq \kappa(L^* - L)/\pi\,W_g$.
At small overlap, $(L^* - L)$, between the electrodes and the gate is can be disregarded.
Moreover, this ac current does not affect  boundary conditions~(22).
,

Considering this and disregarding the nonlinear term in the expression for  the current density (only in the calculation of the linear response), the boundary conditions
can be set as follows:
 
\begin{multline}\label{22}
\delta \varphi_{\omega}|_{x = 0} = \delta V_{\omega},\qquad  \delta \varphi_{\omega}|_{x = L} = 0,\\
\delta \varphi_{\omega}|_{x = L -l -0} = \delta\varphi_{\omega}|_{x = L -l +0},\qquad\\
- \langle \sigma_{\omega}\rangle 
d \delta\varphi_{\omega}/d x|_{x = L -l}= Y_{\omega}\delta\varphi_{\omega}|_{x = L -l}.
\end{multline}
Similar boundary conditions were used previously in the modeling
of different plasmonic devices with the Schottky contacts~\cite{24,25}

Equation~(16) with the boundary conditions (22)  yield the following formulas for the ac potential $\delta\varphi_{\omega}$:

\begin{multline}\label{23}
\frac{\delta\varphi_{\omega}}{\delta V_{\omega}} =\cos(\ae_{\omega}x)\\ 
+ \frac{\xi_{\omega}\sin(\ae_{\omega}{\cal L}) - \cos(\ae_{\omega}{\cal L})}{\xi_{\omega}\cos(\ae_{\omega}{\cal L}) + \sin(\ae_{\omega}{\cal L})}\sin(\ae_{\omega}x)
\end{multline}
for the range $0 \leq x \leq {\cal L}$, where ${\cal L} =(L -l)$ (with $l = L - {\cal L}\ll L$), and

\begin{multline}\label{24}
\frac{\delta\varphi_{\omega}}{ \delta V_{\omega}} =\frac{(L -x)}{l}
\biggl[\cos(\ae_{\omega}{\cal L})\\
+ \frac{\xi_{\omega}\sin(\ae_{\omega}{\cal L}) - \cos(\ae_{\omega}{\cal L})}{\xi_{\omega}\cos(\ae_{\omega}{\cal L}) + \sin(\ae_{\omega}{\cal L})}\sin(\ae_{\omega}{\cal L})
\biggr].
\end{multline}
for ${\cal L} \leq x \leq L$. Here 

\begin{equation}\label{25}
\xi_{\omega} = \frac{\langle\sigma_{\omega}\rangle \ae_{\omega}}{Y_{\omega}}
 =
i\frac{\nu}{\Omega_g}\sqrt{\frac{\omega}{\omega + i\nu}}\frac{1}{(h^{-1} -i\omega\,\tau_c)}.
\end{equation}
where $h^{-1} = 2L/\pi\langle \sigma_{0}\rangle\, r$ and $\tau_c= 2LC/\pi\langle \sigma_{0}\rangle$ is the
time of the Schottky contact "capacitor" charging via the quasi-neutral region of the CNT network.

In particular, for the ac voltage drop across the contact region which determines
the rectified current Eqs.~(23) and (24) yield

\begin{equation}\label{26}
\frac{\delta\varphi_{\omega}|_{x = {\cal L}}}{\delta V_{\omega}} =
\frac{\xi_{\omega}}{\sin(\ae_{\omega}{\cal L}) + \xi_{\omega}\cos(\ae_{\omega}{\cal L})}
\end{equation}

Hence, the 	square of the absolute value of  the ac potential drop across the depletion region averaged over the THz radiation period
$\overline {|\delta V_{\omega}^{DR}|^2} \simeq  
\overline{|\delta\varphi_{\omega}|^2}|_{x = L-l}$ is given by

\begin{equation}\label{27}
\frac{\overline{|\delta V_{\omega}^{DR}|^2}}{(\delta V_{\omega})^2}
= \frac{1}{2}\biggl|\frac{\xi_{\omega}}{\sin(\ae_{\omega}{\cal L}) + \xi_{\omega}\cos(\ae_{\omega}{\cal L})}
\biggr|^2.
\end{equation}

Assuming $m_s \simeq  6\times 10^{-29}$~g, $\kappa = 4$, $\theta = 0.5$,$W_g = 50$~nm, and $\Sigma_{s0} + \overline{\Sigma_{m0}} = 10^{12}$~cm$^{-2}$,  a $L = 0.35-0.70~\mu$m, Eq.~(21) yields
$\Omega_g/2\pi \simeq 0.6 - 1.2$~THz. 
%
The above electron density is achieved
at $V_g \simeq 1.1$~V (i.e., at $V_g/W_g \simeq 4.5\times 10^5$~V/cm).
Considering  these parameters, we also set  $\nu = 2\times10^{12}$~s$^{-1}$, 
$\theta = 0.5$, $R_s = 10^3~k\Omega$
(this corresponds to the experimental data for purely s-CNT network), $N_s = 2\times 10^5$~cm$^{-1}$ (i.e., $r \simeq R_s/N_s = 5~\Omega\cdot$cm). 
As a result,
 we obtain the following estimates: $h\simeq 48 - 96$ and  
 $\tau_c \simeq (0.23 - 0.46)\times 10^{-13}$~s.
 As seen from the results below, just the latter  time determines the role of the Schottky contact capacitance in the high-frequency
response.

\subsection{Ungated structures}

In the CNT structures under consideration with a relatively thick gate layer ($W_g > L$),
the self-consistent electric field is localized near the CNT network plane and sufficiently far from the gate. In these structures, the conductivity of the gate
only weakly affects the plasma oscillations, in particular, their dispersion characteristics. Therefore, such structures 
can approximately be considered as the ungated ones, although  a remote gate can be used for the voltage control of the carrier density. 

In the ungated structures,
the self-consistent ac electric potential $\delta\psi_{\omega} = \delta\psi_{\omega}(x,z)$ in the  channel (z = 0)
and in the areas surrounding it ($z > 0$ and $z < 0)$,
can be found from the following equation~\cite{24,25,52,56,57}:

\begin{equation}\label{28}
\frac{\partial^2 \delta \psi_{\omega}}{\partial x^2} +\frac{\partial^2 \delta \psi_{\omega}}{\partial z^2} = \frac{4\pi\,e\delta \Sigma_{\omega}}{\kappa}\,\delta(z).
\end{equation}
Here the Dirac delta function $\delta (z)$ describes the hole localization
at the plane of the CNT network $z = 0$.
Considering the hole motion along the CNTs, Eq.~(28) can be transformed to the following:

\begin{equation}\label{29}
\frac{\partial^2 \delta \psi_{\omega}}{\partial x^2} +\frac{\partial^2 \delta \psi_{\omega}}{\partial z^2} = \frac{4\pi\langle\sigma_{\omega}\rangle}{i\kappa\omega}
\frac{\partial^2 \delta \psi_{\omega}}{\partial x^2}\,
\delta(z).
\end{equation}

Apart from the natural boundary conditions $\delta \psi_{\omega}|{z = -\infty}
= \delta \psi_{\omega}|{z = -\infty} = 0$, the same boundary conditions as above
[i.e., Eq.~(22)]  can be used for different point in the channel.
Solving Eq.~(29)
 with the latter boundary conditions, for the ac potential distribution in the main part of the channel we obtain

\begin{multline}\label{30}
\frac{\delta\psi_{\omega}}{ \delta V_{\omega}}=
\biggr[\cos(\gamma_{\omega}x)\\ 
+ \frac{\zeta_{\omega}\sin(\gamma_{\omega}{\cal L}) - \cos(\gamma_{\omega}{\cal L})}{\zeta_{\omega}\cos(\gamma_{\omega}{\cal L}) + \sin(\gamma_{\omega}{\cal L})}\sin(\gamma_{\omega}x)\biggr]
\exp[- \gamma_{\omega}|z|],
\end{multline}

\begin{equation}\label{31}
\frac{\delta\psi_{\omega}|_{x = {\cal L}, z = 0}}{ \delta V_{\omega}}=
\frac{\zeta_{\omega}}{\sin(\gamma_{\omega}{\cal L}) +\zeta_{\omega}\cos(\gamma_{\omega}{\cal L})}.
\end{equation}
Consequently,

\begin{equation}\label{32}
\frac{\overline{|\delta V_{\omega}^{DR}|^2}}{(\delta V_{\omega})^2}
=
\frac{1}{2}\biggl|\frac{\zeta_{\omega}}{\sin(\gamma_{\omega}{\cal L}) +\zeta_{\omega}\cos(\gamma_{\omega}{\cal L})}
\biggr|^2.
\end{equation}
One can see that Eq. (32)
coincides  Eqs.~(27) with the replacement of 
 the quantities $\ae_{\omega}$  and  $\xi_{\omega}$ by  
 
\begin{equation}\label{33}
\gamma_{\omega} = \frac{\pi\,\omega(\omega + i\nu)}{2\Omega^2L}.
\end{equation}
and
\begin{equation}\label{34}
\zeta_{\omega} = \frac{\langle \sigma_{\omega}\rangle \gamma_{\omega}}{Y_{\omega}}\\
=i\biggl(\frac{\nu\omega}{\Omega^2}\biggr)\frac{1}{(h^{-1} -i\omega\,\tau_c)},
\end{equation}
respectively.

Here the  pertinent  characteristic plasma frequency is given by the following  equation [different from Eq.(20)]:

\begin{multline}\label{35}
\Omega =  \sqrt{\frac{\pi^2e^2\theta}{L\kappa}\biggl(\frac{\Sigma_0^s}{m_s}
+ \frac{\Sigma_0^m}{m_m}\biggr)}\\
 = \sqrt{\frac{\pi^2e^2\theta(\Sigma_0^s
+ \overline{\Sigma_0^m})}{L\kappa\,m_s}}.
\end{multline}
Hence, $\Omega_g = \Omega\sqrt{\pi\,W_g/L}$, i.e., at the same $L$, $\Omega_g < \Omega$. For the hole density  used in the estimate of $\Omega_g/2\pi$ in the previous subsection, the plasma frequency  $\Omega/2\pi = 0.6 - 1.2$~THz can be achieved in the ungated structures with $L \simeq 0.78 - 3.12~\mu$m, i.e., in the structures with  markedly longer channels. For $\Omega = \Omega_g$, one obtains
$h = h_g$ and $c = c_g$.

As follows from Eq. (32), the quantity $\overline{|\delta V_{\omega}^{DR}|^2}/(\delta V_{\omega})^2$ in the ungated structures (similar to that in the gated structures)  can be much larger than unity at the plasmonic
resonances  $\omega/\Omega = \sqrt{2n-1}
 = 1, \sqrt{3}, \sqrt{5},...$ if the quality factor of the plasma oscillations
 $\Omega/\nu$ (in the ungated structures) is sufficiently large.

If the channel length $L$ and the gate layer thickness $W_g$ are of the same order of magnitude, Eqs.~(27) and (32) are sill valid but the dispersion factors $\ae_{\omega}$ and $\gamma_{\omega}$ should be replaced by the more complex unified factor.

\subsection{Limiting cases}

\begin{figure*}[t]
\centering
\includegraphics[width=12.0cm]{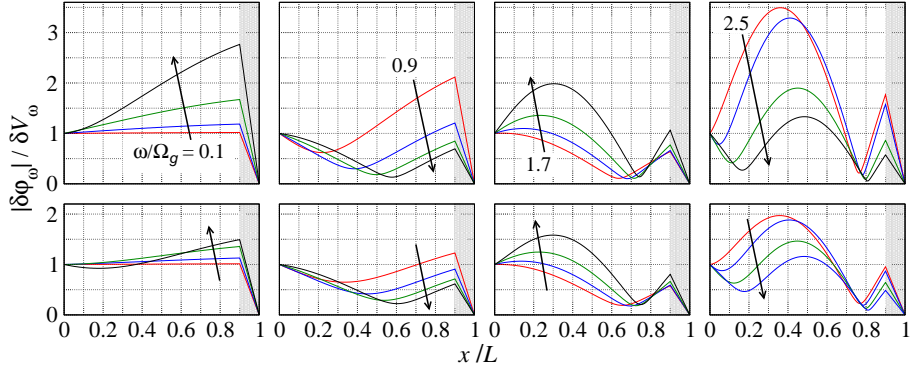}
\caption{Spatial distributions of the ac potential amplitude $|\delta\varphi_{\omega}|/\delta V_{\omega}$ from the Ohmic ($x/L = 0$) to Schottky ($x/L =1$) contacts
calculated for   different ratios $\omega/\Omega_g$ (from $\omega/\Omega_g = 0.1$
to 3.1)  at  $\nu/\Omega_g = 1/\pi$ (upper panels) and  $\nu/\Omega_g = 2/\pi$ (bottom panels). Arrows in each panel show the direction of $\omega/\Omega_g$ increase from indicated values ($\omega/\Omega_g =0.1$, 0.9,  1.7, and 2.5) with 0.2 steps.
The shaded
areas ($0.9 \leq x/L \leq 1$) correspond to the Schottky contact depletion region.
}
\end{figure*}

At low frequencies $\omega \ll \nu$,  $\Omega_g^2/\nu$, $\Omega^2/\nu$, 
when $|\xi_{\omega}|, |\zeta_{\omega}| \ll 1$,
from Eqs.~(27) and (32), we obtain

\begin{equation}\label{36}
\frac{\overline{|\delta V_{\omega}^{DR}|^2}}{(\delta V_{\omega})^2}\simeq
\frac{1}{2}\biggl[1 - \frac{1}{1 + (\langle \sigma_0\rangle r/L)}\biggr]^2.
 \end{equation}
If the contact resistance $r$ is small in comparison with the dc  resistance
of the main part of the CNT network $L/\langle \sigma_0\rangle$ (that can be a large number of the m-CNTs with very small contact resistance)
Eq.(36) yields a very small value:

\begin{equation}\label{37}
\frac{\overline{|\delta V_{\omega}^{DR}|^2}}{(\delta V_{\omega})^2}\simeq
\frac{1}{2}\biggl(\frac{\langle \sigma_0\rangle r}{L}\biggr)^2.
 \end{equation}

In real CNT networks  the contact resistance $r$ can be  larger than
$L/\langle \sigma_{0}\rangle$ as in
 the CNT networks experimentally studied in Ref.~\cite{34} 
 (see, also, Ref.~\cite{51}).
Relatively large contribution to the  resistance of the contacts between quasi-metallic CNTs and the contact electrode (which could form the Schottky contact with rather low energy barrier) can be attributed to quantum contact resistance and the presence of an additional insulating barrier (not shown in Fig.~2) if no special formation methods are used~\cite{58,59}.

In such a case, from Eq.~(36) we obtain

\begin{equation}\label{38}
\frac{\overline{|\delta V_{\omega}^{DR}|^2}}{(\delta V_{\omega})^2}\simeq
\frac{1}{2}\biggl[1 -  \biggl(\frac{L}{\langle \sigma_0\rangle r}\biggr)\biggr]^2,
 \end{equation}
i.e., $ {|\delta V_{\omega}^{DR}|^2}/(\delta\,V_{\omega})^2 \lesssim 1/2$. 

At  relatively large contact resistance, in a wide frequency range  parameter $|\xi_{\omega}| \gg 1$. In this frequency region  for the gated structures Eq.~(27)
yields

\begin{equation}\label{39}
\frac{\overline{|\delta V_{\omega}^{DR}|^2}}{(\delta V_{\omega})^2}
\simeq
\frac{1}{2|\cos(\ae_{\omega}{\cal L})|^2},
\end{equation}

At very high signal frequencies, the imaginary part of the contact region admittance $Y_{\omega}$
can become large, so that  the displacement current  can actually shunt the depletion
region. In this frequency range, $|\xi_{\omega}| 
\propto   1/\sqrt{h^{-2} + \omega^2\tau_c^2}$ (and $\zeta_{\omega} \propto  1/\sqrt{h^{-2} + \omega^2\tau_c^2}$)
can become small. In this case, 

\begin{multline}\label{40}
\frac{\overline{|\delta V_{\omega}^{DR}|^2}}{(\delta V_{\omega})^2}
\simeq
\frac{|\xi_{\omega}|^2}{2|\sin(\ae_{\omega}{\cal L})|^2}\\
 \simeq \frac{1}{2|\sin(\ae_{\omega}{\cal L})|^2} \frac{\omega}{\sqrt{\omega^2 + \nu^2}}\frac{1}{\omega^2\tau_c^2}.
\end{multline}
Equation~(40) explicitly describes a decrease in $\overline{|\delta V_{\omega}^{DR}|^2}/(\delta V_{\omega})^2$ with increasing product $\omega\,\tau_{c}$.
Hence, a decrease in  $\overline{|\delta V_{\omega}^{DR}|^2}/(\delta V_{\omega})^2$
with increasing frequency due to the capacitance effects is characterized by relatively short charging time $\tau_c$. 

The relationships similar to Eqs.~(39) and (40) can be obtained from Eq.~(32) with $\zeta_{\omega}$ and $\gamma_{\omega}$ instead of $\xi_{\omega}$ and $\ae_{\omega}$.

Figure~3 shows the spatial distributions of the ac potential distribution $\delta_{\omega}\varphi = \delta\varphi_{\omega}(x)$
in the gated structures calculated using Eqs.~(23) and (24) for different
ratios $\omega/\Omega_g$ at different ratios $\nu/\Omega_g $ (i.e.,different values of the plasma oscillations  quality factor $\Omega_g/\nu$. It is assumed that $l/L = 0.1$ and $h = 60$ and  $\Omega_g\tau_c = 2/3$. 
If, in particular,$\nu = 2\times10^{12} $~s$^{-}$,the upper and lower panels of Fig.~3 correspond to
$\Omega_g/2\pi = 1.0$~THz and  $0.5$~THz, as well as $\tau_c \simeq 0.3\times10^{-13}$~s and $\tau_c \simeq 0.6\times10^{-13}$~s, respectively

The spatial distributions for $\omega/\Omega_g \sim 1$ (in two left-side upper  pannels)
correspond to the plasma wavelength $\lambda \sim 4L$. 
As seen from Fig.~3 and follows from Eqs.~(23), (24), and (26), 
the absolute value of the ac potential $|\delta \varphi_{\omega}|$ can markedly
exceed $\delta\,V_{\omega}$ at the plasmonic
resonances which are close to $(2n - 1)\Omega_g$, where $n = 1, 2, 3,...$.
In the latter cases the potential drop across the Schottky
contact $|\delta\varphi_{\omega}|_{x = L -l}|$ is large and 
${\overline |\delta V_{\omega}^{DR}|^2}/(\delta V_{\omega})^2 \gg 1$
[see Eq.~(39), in which the denominator reaches a minima  at the plasmonic resonances]. 
the distribution in two left-side panels correspond to the frequency range around
the fundamental plasmonic resonance ($2n - 1=1$), while two right-side panels correspond
to the next resonance ($2n - 1=2$).
Similar situation occurs when in the ungated structures the ratio $\omega/\Omega$ approaches to $\sqrt{2(n-1)}$.

\section{Detector responsivity}

\subsection{Photocurrent}

The variation of the dc current density  due to the THz radiation (THz photocurrent)
$\Delta j_0$,
 which is associated with  the rectified current component  generated by the THz radiation, can be presented as
 
\begin{equation}\label{41}
\Delta j_0=  N_s\beta_s\overline{|\delta\,V^{DR}_{\omega}|^2}.
\end{equation}
with $\overline{|\delta\,V^{DR}_{\omega}|^2}$ given by Eqs.~(27) or (32).

Considering that the current density across the Schottky barriers in the s-CNT-electrode junction is given by $j_s = N_s J_{S}\exp(-e\Phi_S/T)[\exp(e\varphi/T) - 1]$, where $J_S$ is the saturation current of a single-CNT junction and $e\Phi_S$ the height of the Schottky barrier,
one can obtain $\beta_s =  J_{S}J_{S,0}(e^2/2T^2)\exp[-e(\Phi_S - V_0)/T)$.
Taking this into account from Eq.~(42), we obtain

\begin{multline}\label{42}
\Delta j_0= N_s\beta_s\overline{|\delta V_{\omega}^{DR}|^2}\\
=
N_sJ_S\exp\biggl[\frac{e(V_0 - \Phi_S)}{T}\biggr]
 \frac{e^2\overline{|\delta V_{\omega}^{DR}|^2}}{T^2}.
\end{multline}

\subsection{Current responsivity}

Considering the variation of the dc current $D\Delta j_0$ (where $D$ is the device size in the lateral direction perpendicular to the current direction) under the incoming THz radiation of intensity $I_{\omega}$ as the output signal,
the detector current responsivity (in A/W) is defined as 

\begin{equation}\label{43}
{\cal R}_{\omega} = \frac{ED\Delta j_0}{SI_{\omega}},
\end{equation}
where $S = \lambda_{\omega}^2G/4\pi$ is the antenna aperture, $G$ is the antenna gain, and the factor $E$ ($E < 1$ or $E \ll 1$) characterizes the radiation loss
and reflection (effects of mismatching of the antenna and the CNT detector structure).
Taking into account that $(\delta\,V_{\omega})^2 \sim (8\lambda_{\omega}^2/\pi c)I_{\omega}$, where $c$ is the speed of light and $\lambda$ is the wavelength of the incident THz radiation, and using Eqs.~(13), (16), (27), and (32), for the CNT detectors based on the gated and ungated structures
we respectively obtain 

\begin{equation}\label{44}
{\cal R}_{\omega}/{\overline{\cal R}}  = 
 \biggl|\frac{\xi_{\omega}}{\sin(\ae_{\omega}{\cal L}) + \xi_{\omega}\cos(\ae_{\omega}{\cal L})}\biggr|^2,
\end{equation}

\begin{equation}\label{45}
{\cal R}_{\omega}/{\overline{\cal R}} =
\biggl|\frac{\zeta_{\omega}}{\sin(\gamma_{\omega}{\cal L}) +\zeta_{\omega}\cos(\gamma_{\omega}{\cal L})}
\biggr|^2.
\end{equation}
Here 
\begin{multline}\label{46}
{\overline {\cal R}} = \frac{8N^s\beta^sED}{cG}\\
=
\frac{8e^2N^sJ_SED}{cGT^2}\exp\biggl[\frac{e(V_0 - \Phi_S)}{T}\biggr]
\end{multline}
is the responsivity without the plasmonic effects.
Naturally it is proportional to the density of the s-CNTs and depends on the parameters
of their Schottky contact to the metal electrode ($J_S$ and $\Phi_S$).
The maximum of ${\overline {\cal R}}$ is achieved when $\Phi_s -V_0 = 2T/e$, i.e.,
when $\Phi_s - V_0 \sim 50$~meV at room temperature.

The second terms in the right-hand sides of Eqs.~(44) and (45) describe the effect
of the plasma oscillations excitation by the incident THz radiation.
These terms can exhibit a pronounced resonant behavior as a function of the signal frequency $\omega$ and the plasma frequencies $\Omega_g$ (or $\Omega$) provided
the quality factor of the plasma oscillations is large.

Figure~4 shows the spectral characteristics frequency dependences of the normalized current responsivity 
${\cal R}_{\omega}/ {\overline {\cal R}}$ of the gated and ungated detectors calculated using Eqs.~(45) and (46) 
respectively,  
with  different ratios $\nu/\Omega_g$ (upper panel) and $\nu/\Omega$ (lower panel) assuming  $h =60$ and 
$\Omega_g\tau_c = 2/3$ and $\Omega\tau_c = 2/3$.
As seen from Fig.~4, the responsivities of both gated and ungated detectors exhibit resonant maxima with the sharpness increasing with decreasing 
$\nu/\Omega_g$ and $\nu/\Omega$ (with increasing resonance quality factors).
The responsivities can achieve rather high values markedly exceeding the those at low-frequencies. 

As an example, Fig.~5 demonstrates  the frequency dependences  of the gated and ungated detector responsivity with $\Omega_g/2\pi = 1.0$~THz and $\Omega/2\pi = 1.635$~THz,
respectively. These values correspond to $L = 0.42~\mu$m and $W_g = 0.05~\mu$m
(in the gated device), $L = 0.42~\mu$m (in the ungated device),  and  the same hole densities ($\Sigma_0 = 10^{12}$~cm$^{-2}$) in both.
The collision frequency is assumed to be $\nu = 2\times 10^{12}$~s$^{-1}$, so that
$\tau_c \simeq 0.28\times10^{-13}$~s;
other parameters are the same as for Fig.~4.

Both Fig.~4 and Fig.~5 show a marked shift of the resonances from their nominal values $\omega_n = (2n-1)\Omega_g$ and $\omega_n = \sqrt{(2n-1)}\Omega$ toward
smaller frequencies.
Some deviation of the plasmonic resonant frequencies from $(2n - 1)\Omega_g$
and $\sqrt{2(n-1)}\Omega$ is associated with 
the finiteness of its depletion width (${\cal L} \neq L$) resulting in an effective shortening of the plasmonic cavity,   and   
the  plasma oscillations damping due to the hole scattering proportional to $\nu$.
The first reason provides some increase in the resonant frequencies [by a factor of
$(1 + l/L)$], while the second effect  leads to a vary decrease in the resonant frequencies in the ungated devices (see Appendix A). Both mechanisms are relatively weak at the parameters
used in the calculations: $l \ll L$ and  $\nu \ll \Omega_g, \Omega$.
The main contribution to a substantial decrease in the resonant frequencies is 
 associated with the effect of the Schottky contact capacitance.
 
The resonant peak width $\Delta \omega_n$ is  determined by the plasma resonance quality factors,
$\Omega_g/\nu$ and $\Omega/\nu$ [see Eqs.~(A5) and (A6)  in the Appendix A]. 
As for the  height of the resonant peaks with different index $n$,
its decrease with increasing $n$ in the gated devices is solely due to the capacitance effect characterized by the quantity 
 $\omega_n\tau_c$. Our  calculations   shows an equalizing of the peak heights, whose positions tend to the nominal resonant frequencies, when
the charging time $\tau_c$ and the capacitance $C$ decrease. 
 However, in the ungated devices, lowering of the peaks  associated
 with the capacitance effect is reinforced by entirely plasmonic mechanism. This explains a more dramatic responsivity roll-off 
in the ungated devices compared to the gated ones seen in Figs.~4 and 5.

\begin{figure}[t]
\centering
\includegraphics[width=6.0cm]{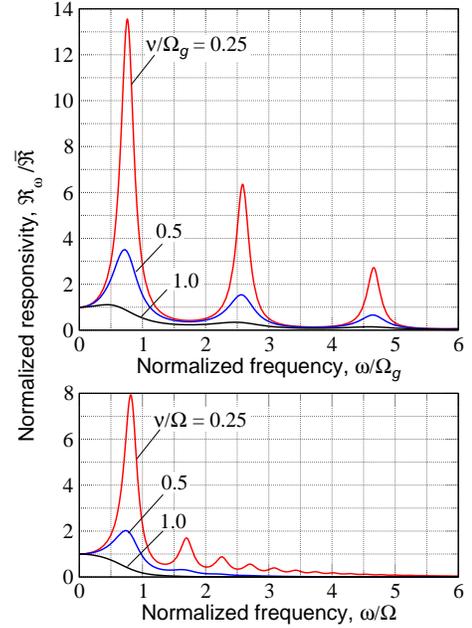}
\caption{Spectral dependences of the normalized  current responsivity 
${\cal R}_{\omega}/ {\overline {\cal R}}$ of  gated detectors (upper panel) for different ratios of 
 $\nu/\Omega_g$ and  ungated detectors (lower panel) for different ratios of 
 $\nu/\Omega$.
}
\end{figure}

\begin{figure}[t]
\centering
\includegraphics[width=6.0cm]{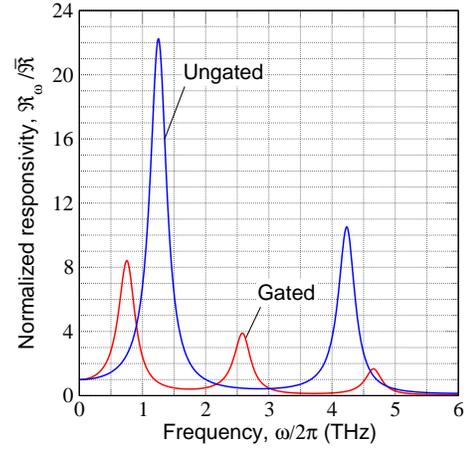}
\caption{Normalized  current responsivity 
${\cal R}_{\omega}/ {\overline {\cal R}}$ versus signal frequency $\omega/2\pi$
for  gated  and ungated  detectors 
with  $\Omega_g/2\pi = 1.0$~THz and $\Omega/2\pi = 1.635$~THz, respectively, and $\nu = 2\times 10^{12}$~s$^{-1}$.
}
\end{figure}

\subsection{Voltage responsivity}

In the absence of the applied bias voltage, the quantity $V_0$ can be considered as the induced photovoltage. The latter is determined
by the balance between the photocurrent  $D\Delta j_0$ and the dc current density  caused by the induced photovoltage $j_0 = V_0(N_s/R_s + N_m/R_m)$.
As a result, using Eq.~(43), assuming that $V_0 \ll T/e$ (i.e., at relatively low
radiation intensities),  for the voltage responsivity ${\cal R}_{\omega}^V$
we find

\begin{equation}\label{48}
{\cal R}_{\omega}^V = {\cal R}_{\omega}\frac{\rho}{D},
\end{equation}
where

\begin{equation}\label{49}
\rho = \frac{R_sR_m}{(N_mR_s + N_sR_m)}
\end{equation}

From Eq.~(48)  we obtain that in the structures with not too small fraction of the m-CNTs in which $N_m/N_s > R_m/R_s$
$R_{\omega}^V \simeq {\cal R}_{\omega}
(R_m/N_mD)$ can be rather small.
In contrast, in the structures with the destructed m-CNTs or disconnected from the right contact (the Schottky contact for the s-CNTs) by the scorching (burning)  of the near contact portions of the m-CNTs $R_{\omega}^V$ can be rather large.
Indeed, setting $N_m = 0$, from Eq.~(48) we obtain
$R_{\omega}^V \simeq {\cal R}_{\omega}(R_s/N_sD)$.
One needs also to point out that in the former case, the current responsivity
${\cal R}_{\omega}$ can be much smaller  than in the latter case. 
Thus, the elimination of the m-CNTs can lead to
a marked increase in the low-frequency current responsivity but even stronger
relative increase in the low-frequency voltage responsivity. 
In principle, the elimination of the m-CNTs can result in a significant change of the plasmonic properties.

\section{Comparison with the experimental data}

\begin{figure}[t]
\centering
\includegraphics[width=5.0cm]{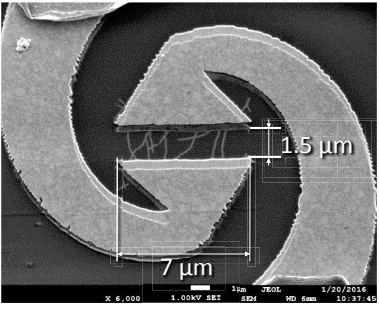}\\
\caption{SEM image of a fragment of the device structure with a spiral antenna.
}
\end{figure}

\begin{figure}[t]
\centering
\includegraphics[width=6.0cm]{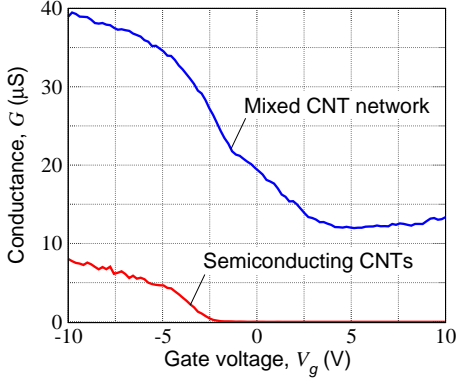}
\caption{ Transfer characteristics 
(conductance vs gate voltage)  of original mixed CNT structure
and CNT structure after selective breakdown of m-CNTs.
}
\end{figure}

We compare the above theoretical results  with the recent experimental data
related to the 
THz detectors based on  asymmetric devices with one the Ohmic and the Schottky contacts and   the conduction channels are
formed by lateral arrays of CNTs ~\cite{34} and some new our experimental results 
shown below.
Assuming $J_S = (2 - 6)\times10^{-7}$~A,  $e(\Phi_s - V_0)/T = 2$, and $T = 300$~K,
we obtain $\beta_s \simeq (2 - 6)\times 10^{-5}$~A/V$^2$, that is consistent with
the experimental value ~\cite{34} $\beta^s \simeq 4\times 10^{-5}$A/V$^2$. For $ND = (N_s + N_m)D = 100$ ($N_s/N_m = 2$), and $G = 1.5$, Eq.~(46) yields ${\overline {\cal R}} 
\simeq (0.2-0.6)E$~A/W. Taking into account  that in the experiments~\cite{34}
$E \sim 10^{-3}$ and  setting also $\rho/D = 10~k\Omega$, we find ${\cal R}_{0}\simeq (3-6)\times 10^{-4}$~A/W and somewhat lower values for $\omega/2\pi = 140$~GHz (that is markedly smaller than the plasmonic frequency).
Taking into account that $R^V_0 \simeq \overline{{\cal R}}\,(R_s/N_sD)$, for the sub-THz frequency range 
we obtain  
 $R_{\omega}^V \lesssim R^V_0  \simeq (3 - 6)$~V/W. These values of the current and voltage responsivities are  in a reasonable agreement with the data obtained experimentally~\cite{34}.  
To achieve higher responsivities, the devices with higher antenna efficiency
should be used. Another opportunity is to realize the resonant plasmonic response
predicted above 
using the structures with relatively high ratios $\Omega_g/\nu$ or $\Omega/\nu$.

As predicted above, the elimination of the m-CNTs might also lead to a substantial 
increase in the responsivity. To verify this, we fabricated and measured 
device  structures similar to those  fabricated and studied as in Ref.~\cite{34}.
The spacing between the contacts and the gate thickness are $L = 1500$~nm and $W_g = 500$~nm, respectively. It  includes eight CNTs connecting the electrodes (so that $N \simeq  10^4$~cm$^{-1}$) and was supplied with a spiral antenna.
The SEM image of such a CNT  device is shown in Fig.~6. The structure corresponds to that  schematically shown in Fig.~1.
Originally the CNT network was formed with a mixture of the s- and m-CNTs.
This is confirmed by the analysis of the device transfer characteristics. 
The response of the structure was measured at the negative gate voltage $V_g = - 10$~V) as a function of the radiation frequency in the sub-THz range from 140 to
220 GHz The CNT network was coupled to the radiation by a spiral antenna.
Then  the m-CNTs that provided a substantial current path
between the electrodes were destroyed, so that the current between the electrodes was associated with the s-CNTs. 
The response of this structure after the destruction of of the m-CNTs was measured as a function of the radiation frequency
(at the same gate voltage and in the same frequency range as in the case of original structure). The results of the transfer characteristics measurements are shown in Fig.~7. As seen from Fig.~7, the maximum device conductance (ON-conductance) at the negative gate voltage originally was about of $40~\mu$S. After some CNTs were destroyed, the conductance droped down to $8~\mu$S. This implies that the contribution of the destroyed m-CNTs was about of $32~\mu$S, and that
the hole density in the s-CNTs was about one fifth of the net density. The minimum conductance (OFF-conductance) of the original structure was 
about of $12.5~\mu$S.

As seen from Fig.~8, the voltage responsivity of the device with the destructed m-CNTs in the frequency range 150 - 160~ GHz is much larger (about five times) than that of the original device with a mixture of the m- and s-CNTs. 

\begin{figure}[t]
\centering
\includegraphics[width=6.0cm]{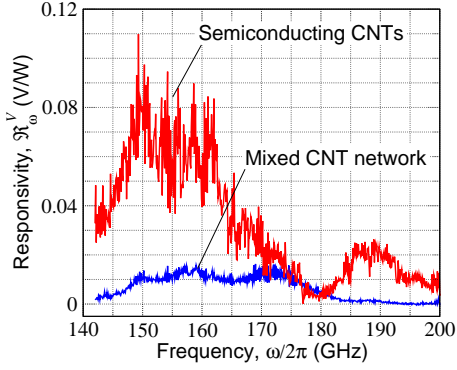}
\caption{Voltage responsivity ${\cal R}_{\omega}^V$ versus radiation frequency $\omega/2\pi$ before (mixed CNT network) and after m-CNT destruction (s-CNT network) at $V_g = -10$~V.
}
\end{figure}

\begin{figure}[t]
\centering
\includegraphics[width=6.0cm]{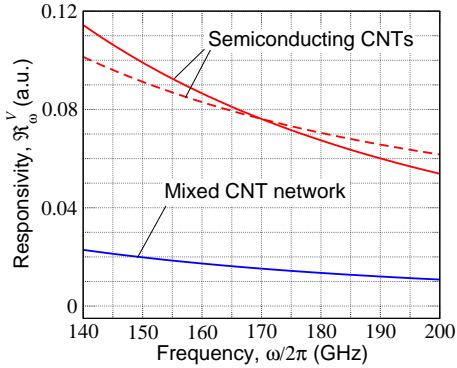}
\caption{Voltage responsivities ${\cal R}^V_{\omega}$ versus frequency $\omega/2\pi$ calculated  
 for low-density   CNT detectors with a CNT mixture 
 and with destructed m-CNTs:  disconnected (solid line) and  fragmented 
 (dashed line) m-CHTs.
}
\end{figure}

First of all, our model describes the mechanism of the THz response of the
devices under consideration in terms of the rectification of the signals
at the Schottky contacts of the s-CNTs and the electrode. As shown, the shunting effect caused by relatively low contact resistance of m-CNTs can substantially 
suppress the effect of rectification. The elimination (disconnection from the electrode or destruction) of the CNTs lead to an efficient increase in the responsivity. 
This is confirmed by the data shown in Fig.~8.

Second, the  device model developed above predicts the resonant response of the
CNT devices under consideration to the THz radiation associated with the excitation of the 2D-plasmons. 

Although the experimental data shown in Figs.~7 and 8 are related to rather delute CNT networks, we apply the obtained above formulas for, at least, qualitative 
consideration of the experimental situation.

Since the gate layer thickness $W_g = 500$~nm is close
the quantity $L/\pi \simeq 477$~nm, $\Omega_g \sim \Omega$, so that both Eqs.~(21)
and (36) can be used for rough estimates.
There are two plausible  scenario: (a) due to burning, the m-CNTs are disconnected from
the electrode, which forms the Schottky contacts with the s-CNTs,  and (b)
the m- CNTs  are fragmented into small fractions much shorter than the spacing between the electrodes.

In the case "a", the holes belonging to the both types of the CNTs contribute to
the plasma frequencies $\Omega_g$ and $\Omega$. 
The analysis of the measurement results provides the  hole  density induced by
the gate voltage $V_g = - 10$~V
about of $10^{11}$~cm$^{-2}$. The acceptor-induced hole density could be roughly estimated from the threshold gate voltage ($\sim - 2V$, see Fig.~6) to be about 
$2\times 10^{10}$~cm$^{-2}$.
 Setting $\kappa \simeq 4$,
$m_s = (5-6)\times 10^{-29}$~g, $\theta = 0.5$,
$V_g = - 10$~V, 
and $\Sigma_0  = 1.2\times 10^{11}$~cm$^{-2}$, from Eqs.~(20) and (35)
we obtain $\Omega_g/2\pi \simeq \Omega/2\pi  \simeq 316 - 347$~GHz.
In contrast, in the case "b", when only the holes in the s-CNTs determine the plasma frequencies, assuming $\Sigma_0 = \Sigma_{s,0}  \simeq 2.4\times 10^{10}$~cm$^{-2}$, we arrive at
$\Omega_g/2\pi \simeq \Omega/2\pi  \simeq 141 -155$~GHz.

We calculated the current and voltage responsivities of the detectors with the above parameters in a wide range of hole collision frequencies $\nu$.
Figure~9 shows an example of the voltage responsivity ${\cal R}_{\omega}^V$ as a function of the signal frequency $\Omega/2\pi$ calculated using Eqs.~(44) - (48)
for the CNT structures with a mixture of the m- and s-CNTs with  $\Omega_g/2\pi = 330$~GHz and for the CNT structure with disconnected m-CNTs with $\Omega_g/2\pi = 330$~GHz) as well as   fragmented  m-CNTs with $\Omega_g/2\pi = 150$~GHz  assuming the hole collision frequency $nu = 2\times 10^{12}$~s$^{-1}$ and the pertinent charging time $\tau_c = 4\times 10^{-2}$~s .
In line with Fig.~7, it is also assumed that
the ratio of the conductance  $G \propto \rho^{-1}$ for the structure with the CNT mixture and for the structure with destroyed m-CNTs is equal to five.
According to   Fig.~9 (as well as according to the calculations with other values of $nu$), the frequency dependences of the voltage responsivity are monotonically decreasing not exhibiting any resonant peaks. This can be attributed
to relatively low quality factor ($\Omega_g/\nu \simeq 0.47$ in the case of fragmented m-CNTs) and to the signal frequencies well below the plasmonic resonance
frequency in the case of the mixed CNT array and the array with disconected m-CNTs).
Due to this,  t the responsivity maxima in Fig.8 are  most probably  attributed to  to  antenna shape -induced effects.


\section{Discussion}

\subsection{Self-consistent 2D model}
 As it was assumed in the model under consideration,
each electrons interact with a large number of others. This is in line with the consideration of the collective effects in different plasmonic media with
a large  number of charged particles in the Debye sphere~\cite{60}.

In our case,  assuming for simplicity that the Debye screening length $l_D$
is the same as in the standard 2DEG with effective (or fictituous) mass $m = 6\times 10^{-29}$~g, $\kappa = 4 -20$, and $\Sigma_0 = 10^{12}$~cm$^{-2}$,
for the number of the electron inside the region with the area $\pi l_D^2$ (inside the "Debye circle") ${\cal N} = \pi\, l_D^2\Sigma_0 =\pi(\kappa\hbar^2/me^2)^2\Sigma_0$
 we find
${\cal N} = \pi\, l_D^2\Sigma_0$. Setting $\Sigma_0 = 10^{12}$~cm$^{-2}$,
we obtain  ${\cal N}\simeq 1 - 7$.
However, one needs to point out that 
the effects of the self-consistent electric field can really be important in different plasmas even if the parameter of ideality ${\cal N} \lesssim  1$.  

 The value $\Sigma_0 = 10^{12}$~cm$^{-2}$ (which is determined either by the gate voltage and doping level) can
correspond to
  $N = N_s + N_m
\simeq (1 -2)\times 10^5$~cm$^{-1}$.
The quantity $N \simeq(1 - 2)\times 10^5$~cm$^{-1}$ implies that  the average
 distance between CNTs is about of $N^{-1} \simeq 50 - 100$~nm, which is orders of magnitude larger that
 the size of the electron localization around the CNTs (slightly larger than the diameter the CNTs under consideration $2R = 1.5-2.3$~nm), so that the overlap of the electron wave functions of individual CNTs is negligible.
As was mentioned above, the crossing of the CNTs, at which the CNTs touch each other, can play the role of the electron scattering (collision) points. Due to the deformation mechanism of such scattering the effective collision frequency is smaller than the frequency of the crossing of these points by electrons.
For $N \lesssim (1 -2)\times 10^{5}$~cm$^{-1}$ and  the average electron velocity along the CNTs
$\langle v \rangle \simeq (1-5)\times 10^{7}$~cm/s,
this yields for  the pertinent collision frequency the following estimate: 
$\nu_{cross} < \langle v \rangle N \simeq  (0.5 - 5)\times 10^{12}$~s$^{-1}$.
The interaction of the electrons belonging to different CNTs near the crossing of the latter, does not change the momentum of the pairs of colliding electrons, and, hence, does not contribute to the electron momentum relaxation and the quantity $\nu$.
The latter implies that the "crossing" collision mechanisms in question should  not suppress the plasmonic resonant response  when
$\Omega_g/2\pi, \Omega/2\pi > 0.3 - 1.5$~THz. 

\subsection{Bolometric mechanism}
The absorption of the THz radiation in the CNT network and similar sistemscan lead to a heating of
the hole system in the CNTs (see, for example, Refs.~\cite{61,62,63,64}. 
The excitation of the plasmonic oscillation can
result in the enhancement of the absorption at the plasmonic resonances.
In this case, apart from the rectified current, an extra dc current associated with
the difference of the electron temperature in the CNT network and the contacts
can appear if the contacts have different thermal properties. 
The calculation of the bolometric contribution to the responsivity under the conditions of the plasmonic effects, requires a more complex model than that used above. The spatial distribution of the effective hole temperature might be essentially nonuniform due to the nonuniform spatial distribution of the ac electric field
(which heats the DHG in the CNT network under consideration) and due to thermal flows from the CNT network toward the contacts. Therefore, 
the equations of more strict  device model should be supplemented by the heat conductivity equation and related boundary conditions accounting for the contact properties. This model will be considered elsewhere. 
Below we estimate the bolometric contribution to the responsivity using
a fairly simple model.

Assuming that (1) the variation of the current due to the hole heating is associated
primarily with the variation of the Schottky contact electrical conductivity (due to higher thermal conductivity of the Ohmic contact and, hence, smaller temperature
gradient near the latter),  (2) the variation of the effective hole temperature near the Schottky contact
is close to that averaged over the entirely CNT network and the plasmonic oscillation period 
$\overline{\langle \delta T_{\omega}\rangle} > 0$, and (3) only the s-CNTs contribute to the effect in question. In such a situation, 
the expression for the density of the rectified current given by Eqs.~(41) and (42)
should be generalized as follows:

\begin{equation}\label{50}
\Delta j_0 = N_s(\beta_s \overline{|\delta\,V^{DR}_{\omega}|^2} + \gamma_s\overline{\langle \delta T_{\omega}\rangle}),
\end{equation}
where $\gamma_s = [eJ_S(\Phi_S - V_0)/T^2]\exp[-e(\Phi_S - V_0)/T] = 
2\beta_s(\Phi_S - V_0)/e$ characterized the variation of the current density trough the Schottky contact due to an increase of the hole effective temperature in the s-CNTs. Since  $\overline{\langle \delta T_{\omega}\rangle}$
is determined by the balance between   the THz power absorbed by the CNT network
and the power which goes to the lattice (and the contact), one can write  
$\overline{\langle \delta T_{\omega}\rangle} \simeq {\rm Re}\langle\sigma_{\omega}\rangle\overline{|\delta\,V^{DR}_{\omega}|^2}\tau_{\varepsilon}/\Sigma_0L^2$, where $\tau_{\varepsilon}$ is the effective energy relaxation time.
Considering this, one can arrive at the the following formulas for the rectified current density $\Delta j_0$ and the net responsivity ${\cal R}_{\omega}^{net}$
(associated with both the rectification and bolometric effects):

\begin{equation}\label{51}
\Delta j_0 = N_s\beta_s \overline{|\delta\,V^{DR}_{\omega}|^2}(1 + \langle H_{\omega}\rangle)
\end{equation}
and
\begin{equation}\label{52}
{\cal R}_{\omega}^{net}
= {\cal R}_{\omega}(1 + \langle H_{\omega}\rangle).
\end{equation}
Here 
\begin{equation}\label{52}
\langle H_{\omega}\rangle \simeq   \frac{2(\Phi_S - V_0)\tau_{\varepsilon}{\rm Re}\langle \sigma_{\omega}\rangle}{e\Sigma_0L_g^2} =
2\nu\tau_{\varepsilon}b_{\omega}
\end{equation}
is the ratio of the hole  bolometric response and
the effect of the rectification, 
where $b_{\omega} = e(\Phi_S - V_0)/m_sL_g^2(\omega^2 + \nu^2) \simeq e\Phi_S/m_sL_g^2\omega^2 $.
 Setting $e(\Phi_S - V_0) = 50$~meV, $m_s = 6\times 10^{-29}$~g, and 
$L_g = 1~\mu$m, $\nu = 2\times 10^{12}$~s$^{-1}$ at $\omega/2\pi = 0.1 - 1.0 $~THz, we obtain $b_{\omega} \simeq 3\times 10^{-3} - 6\times 10^{-2}$. This implies that
$\langle H_{\omega}\rangle$ can be comparable with unity or exceed it if
the hole energy relaxation time $\tau_{\varepsilon}$ is sufficiently long. 
According to these  estimates, there should $\tau_{\varepsilon} > 10^{-11}$~s
for the sub-THz frequencies and $\tau_{\varepsilon} > 10^{-10}$~s for the frequencies about 1 THz. At sufficiently low temperatures,
the latter conditions can be satisfied.

If the hole heating leads to stronger  hole current from the CNT network to the Ohmic contact due to, in particular, special thermal isolation of the pertinent
CNT network (as in device A in Ref.~\cite{34}),  the bolometric responsivity
can be relatively high and exhibit  the opposite sign.

\section{Conclusions}

 We investigated  the hole transport and the plasmonic oscillations
 excited by incoming THz radiation 
  in the device structures with
 lateral CNT networks and asymmetric contacts (one Ohmic contact 
 and one Schottly contact) and evaluated their effect on the operation of the THz detectors using such structures.
  We  developed the device models accounting  for the 2D nature of the collective behavior of the hole  system in lateral CNT networks, i.e., treating this system as a 2DHG,  including both the s- and m-CNTs and the specifics of the contact phenomena.
  As demonstrated, the rectification of the ac current stimulated by the THz radiation at the Schottky contact of the s-CNTs  (associated with the nonlinearity of the current-voltage characteristics of this contact) is an effective mechanism of the THz detection. The excitation
  of plasmonic oscillations of the self-consistent electric field by the THz radiation can lead to a substantial increase in the rectified current and, hence,
 to the  elevated  detector responsivity. Due to the resonant character
  of the plasmons excitation, the responsivity can exhibit sharp resonant peaks.
  The positions of these peaks can be controlled by the gate voltage. 
  We showed that the disconnection (or fragmentation) of the m-CNTs from the contact electrode
  results in a marked increase in the responsivity due to the elimination of the partial shunting of the Schottky contact. This was confirmed experimentally by comparing the CNT structures with natural ratio of the s- and m-CNTs densities (usually two to one)  with the structures with the destructed m-CNTs.
The  resonant THz CNT detectors with sufficiently high density s-CNT networks
can surpass other THz detectors using plasmonic effects at room temperatures.

\section*{Acknowledgments}
One of the authors (V.R.) is grateful to Dr. N. Ryabova for assistance.
The work was supported by 
the Japan Society for Promotion of Science (Grant-in-Aid for Specially Promoted Research $\#$ 23000008) and  
 the Russian Scientific Foundation (Projects $\#$14-29-00277 and $\#$14-19-01308). G.F., I.G., and G.G.
 acknowledge support of the Russian Foundation for Basic Research (Grants $\#$ 15-02-07787 and $\#$15-02-07841). 
The works at  RPI was  supported by the US Army Research Laboratory
Cooperative Research Agreement.
  
 \section*{Appendix A. Lowering and smearing of resonant plasmonic peaks}
\setcounter{equation}{0}
\renewcommand{\theequation} {A\arabic{equation}}

In a wide range of frequencies the responsivities of the gated in ungated detectors
are

\begin{multline}\label{A1}
{\cal R}_{\omega} \propto |\cos(\ae_{\omega}{\cal L})|^{-2}
= \biggl|\cos\biggl(\frac{\pi\sqrt{\omega(\omega + i\nu)}{\cal L}}{2\Omega_g\,L}\biggr)\biggr|^{-2}\\
\simeq \biggl[\cos^2\biggl(\frac{\pi\omega{\cal L}}{2\Omega_gL}\biggr)
\cosh^2\biggl(\frac{\pi\nu{\cal L}}{4\Omega_gL}\biggr)\\
+
\sin^2\biggl(\frac{\pi\omega{\cal L}}{2\Omega_gL}\biggr)
\sinh^2\biggl(\frac{\pi\nu{\cal L}}{4\Omega_gL}\biggr)
\biggr]^{-1}
\end{multline}
and  
\begin{multline}\label{A2}
{\cal R}_{\omega} \propto |\cos(\gamma_{\omega}{\cal L})|^{-2} 
= \biggl|\cos\biggl(\frac{\pi\omega(\omega + i \nu){\cal L}}{2\Omega^2L}\biggr)\biggr|^{-2}\\
=\biggl[\cos^2\biggl(\frac{\pi\omega^2{\cal L}}{2\Omega^2L}\biggr)
\cosh^2\biggl(\frac{\pi\nu\omega{\cal L}}{4\Omega^2L}\biggr)\\
+
\sin^2\biggl(\frac{\pi\omega^2{\cal L}}{2\Omega^2L}
\biggr)
\sinh^2\biggl(\frac{\pi\nu\omega{\cal L}}{4\Omega^2L}\biggr)
\biggr]^{-1}, 
\end{multline}
respectively

Maxima of functions (A1) and (A2) are achieved at the resonant frequencies
(if $\nu \ll \Omega_g, \Omega$)

\begin{equation}\label{A3}
\omega_n 
\simeq (2n-1)\Omega_g\biggl(1 + \frac{l}{L} \biggr)
\end{equation}
and, neglecting the term $\nu/4\pi\Omega(2n-1)^{3/2} \ll 1$,
\begin{equation}\label{A4}
\omega_n \simeq \sqrt{(2n-1)}\Omega\biggl(1 + \frac{l}{L}\biggr).
\end{equation}
For the broadening of the resonant peaks from Eqs.~(A1) and (A2) we obtain:
at
\begin{equation}\label{A5}
\Delta\omega_n \simeq \frac{\nu}{2}\frac{\omega_n}{\Omega_g} = \frac{(2n - 1)}{2}\nu
\end{equation}
and
\begin{equation}\label{A6}
\Delta\omega_n \simeq  \frac{(2n - 1)}{4}\nu.
\end{equation}

As follows from Eq.~(A1), the  resonant peaks height in the gated detectors 
is given by
\begin{equation}\label{A7}
{\cal R}_{\omega_n} \propto \sinh^{-2} \biggl(\frac{\pi\nu{\cal L}}{4\Omega_gL}\biggr) \simeq \biggl(\frac{4\Omega_g}{\pi\nu}\biggr)^2.
\end{equation} 
At $\omega_n\tau_c \gtrsim 1$, ${\cal R}_{\omega_n}$ becomes proportional to a factor $\omega_n\tau_c)^{-2}$, so that ${\cal R}_{\omega_n}$
drops as $(\omega_n\tau_c)^{-2}\propto (2n - 1)^{-2}$. This implies that  the peaks roll-off 
in such devices is  associated primarily with the capacitance effect characterized by
the charging time $\tau_c$.
In the case of ungated devices, Eq.~(A2) yields

\begin{multline}\label{A8}
{\cal R}_{\omega_n} \propto \sinh^{-2}\biggl(\frac{\pi\sqrt{(2n-1)}\nu}{4\Omega}\biggr)\\ 
\simeq\frac{1}{(2n-1)}\biggl(\frac{4\Omega}{\pi\nu}\biggr)^2.
\end{multline}
Hence, in the ungated devices, ${\cal R}_{\omega_n}$ decreases with the index $n$ starting from its moderate  values.
At $\omega_n\tau_c \gtrsim 1$, ${\cal R}_{\omega_n}$, the capacitance effects lead
to an additional roll-off of 
the responsivity peaks height with increasing $n$.

\end{document}